\documentclass[aip,jcp,amsmath,amssymb,reprint,floatfix]{revtex4-1}

\usepackage{graphicx}
\usepackage{dcolumn}
\usepackage{bm}
\usepackage{color}
\usepackage{hyperref}
\usepackage{multirow}

\newcolumntype{d}[1]{D{.}{.}{#1}}

\usepackage[utf8]{inputenc}
\usepackage[T1]{fontenc}
\usepackage{mathptmx}

\begin{document}

\preprint{AIP/123-QED}

\title{Determination of fundamental properties of nitrogen from first principles. \\ II. Potential energy curve and spectroscopic properties of N$_2$}

\author{Jakub Lang}
\author{Micha\l\ Przybytek}
\author{Micha\l\ Lesiuk}
\email{e-mail: m.lesiuk@uw.edu.pl}
\affiliation{\sl University of Warsaw, Faculty of Chemistry, Pasteura 1, 02-093 Warsaw, Poland}
\date{\today}

\begin{abstract}
This work is the second part of the series devoted to first-principles determination of the fundamental properties of nitrogen. In this part, we calculate the potential energy curve for the electronic ground state of N$_2$. The potential is divided into three components: short-, medium-, and long-range, and a different computational protocol is applied to each component, based on a composite coupled cluster approach, genuine multireference methods, and asymptotic expansion of the interaction energy. A particular focus is on the short-range part, as the accuracy of this component is critical from the point of view of metrological applications, including the temperature dependence of many properties of nitrogen gas. Uncertainties of the theoretical data, originating both from basis set incompleteness and applied theoretical method, are rigorously analyzed and provided at $2\sigma$ level, i.e. expanded ($k=2$) uncertainties. The developed potential energy curve is used to calculate the spectroscopic parameters of the N$_2$ ground state and the results are compared with the available theoretical and experimental data.
\end{abstract}

\maketitle

\section{Introduction}
\label{sec:intro}

In the previous paper of the series (referred to as Paper~I in the following), we outlined the need for reliable theoretical data for the fundamental properties of the nitrogen molecule. In particular, the focus is on quantities such as the polarizability and magnetic susceptibility, which are of fundamental importance in modern thermometry and pressure metrology experiments. \cite{gaiser15,gaiser17,gaiser18,gaiser20,gaiser21,gaiser22,jousten17,gao17,rourke19,ripa21,rourke21} The determination of these quantities for diatomic systems is complicated by the fact that, unlike for noble gas atoms, they depend on the temperature. As direct measurements for a wide temperature range are laborious, theoretical results can supplement the measured quantities and make it possible to transfer the experimental results from one temperature to another. On the theoretical side, temperature dependence of polarizability, as well as other quantities, can be taken into account by using two methods: direct rovibrational averaging within the canonical ensemble or path integral Monte Carlo approach. \cite{Feynman65,Ceperley95,Garberoglio24} Regardless of which method is used, the necessary component is the potential energy curve of N$_2$, i.e. interaction potential between two nitrogen atoms. Therefore, the main objective of this Paper~II is to determine the theoretical potential energy curve of N$_2$ from first principles and provide systematic uncertainty estimates of the calculated results. Our focus is primarily on providing as accurate a representation of the potential as possible with the current theoretical framework in the vicinity of the equilibrium distance. This is because these regions of the potential contribute the vast majority of the temperature dependence and are thus critical from the point of view of metrological applications described above and in Paper~I. While regions of the potential energy curve corresponding to medium and large internuclear separations are also necessary to find the rovibrational levels of the molecule, they can be described by a somewhat less accurate method without overall degradation of the quality of the results relevant from the experimental point of view.

The determination of the potential energy curve of N$_2$ is complicated by the fact that the separation of two nitrogen atoms involves the breaking of a triple bond. Therefore, for medium and large interatomic distances, the wavefunction of the molecule acquires a strong multireference character, which is a well-known challenge for electronic structure methods. For this reason, we split the potential energy curve into three components: short-, medium-, and long-range. In the short-range, i.e., in the vicinity of the equilibrium bond length, the system has only mild multireference character and can be determined using the conventional hierarchy of coupled cluster methods within the Born--Oppenheimer approximation. The leading-order relativistic, quantum electrodynamics (QED), and finite nuclear mass corrections are calculated and included if deemed significant. The overall uncertainty originating from the finite basis set effects, incomplete wavefunction model, etc., is rigorously determined using error propagation rules. In the long-range part of the potential, the interaction energy can be expanded into asymptotic series involving the asymptotic dispersion coefficients, $C_n$, which have been calculated in Paper~I. We analyze the errors of this expansion and use it only for sufficiently large interatomic distances to make the overall uncertainty under tight control. The medium part of the potential is determined using a genuine multireference approach based on the configuration interaction method. However, because the uncertainty of this method is difficult to assess directly, it is estimated by comparison with reliable data from the short- and long-range parts of the potential where the uncertainties are known. By combining the three components of the potential, we determine an analytic representation of the total potential and its uncertainty, which will be used in the next paper in the series.

From the point of view of the aforementioned applications, it is critical that the reported theoretical results carry an uncertainty estimate. Throughout this work, in the extrapolation of the calculated quantities to the complete basis set (CBS) limit and their uncertainty estimation we employ the same scheme as in Paper~I, i.e., Riemann CBS limit formula \cite{lesiuk19Riemann} combined with random walk procedure \cite{Lang25Random} to determine the corresponding error bars. The uncertainty estimates are always given at $2\sigma$ confidence level, also referred to as $k=2$ expanded uncertainties, unless explicitly stated otherwise. Deviations from this protocol are explicitly stated and explained individually in each case.

The following conversion factors are used throughout the present work: $1$~hartree = $27.211\,386$~eV = $219\,474.63$~cm$^{-1}$, and $1$~bohr = $0.529\,177\,2$~\AA, according to the most recent 2022 CODATA recommendations. \cite{mohr25} All distances, unless stated otherwise, are given in bohr.

\section{Short-range potential energy curve}
\label{sec:srpotential}

\subsection{General considerations}
\label{sec:general}

The interaction energies $E_{\mathrm{int}}(R)$ of N$_2$ as a function of the interatomic distance $R$ are calculated as
\begin{align}
\label{eint}
    E_{\mathrm{int}}(R) = 2E_{\mathrm{N}}(R) - E_{\mathrm{N}_2}(R),
\end{align}
where $E_{\mathrm{N}_2}(R)$ is the total energy of the dimer, and $E_{\mathrm{N}}(R)$ is the energy of a single atom. Throughout this work, the Boys--Bernardi counterpoise correction \cite{Boys70} is applied to account for the basis set superposition error, regardless of the electronic structure method used. As a result, both $E_{\mathrm{N}_2}(R)$ and $E_{\mathrm{N}}(R)$ are calculated in the same basis set, i.e., the atomic energy is always evaluated in the two-center basis of the dimer and hence $E_{\mathrm{N}}(R)$ carries a dependence on the internuclear distance. Note that the sign of Eq.~(\ref{eint}) is arbitrarily chosen such that $E_{\mathrm{int}}(R)$ is positive in the regions of the potential where the interaction is attractive.

\subsection{Outline of the composite scheme}
\label{sec:short}

Similarly as in Paper~I we define a composite scheme that enables us to calculate the necessary quantities with controlled precision by combining several electronic structure methods of different quality. However, the composite scheme used for the potential energy curve of N$_2$ is more extensive and multilayered. This is mostly the consequence of the fact that in the nitrogen atom we deal with the dominant contribution coming from the five valence electrons and it is still possible to treat it with the full configuration interaction (FCI) method. \cite{CI_method_1999} For the nitrogen molecule we deal with ten outer-shell electrons and hence the leading valence contribution to the interaction energy has to be split into components calculated at successively higher levels of CC theory. \cite{vcivzek1966correlation,vcivzek1969use,crawford00,bartlett07} In the following, we use the standard abbreviations for the CC hierarchy of methods, namely CCSD denotes the CC variant with single and double substitutions, \cite{purvis1982full,scuseria1987closed} CCSDT denotes the variant with single, double, and triple substitutions, \cite{noga1987full,scuseria1988new} and so on.

The total scheme used in this work to calculate the short-range part of the potential reads
\begin{align}
\label{composite}
\begin{split}
    E_{\mathrm{int}}(R) &= E_{\mathrm{int}}^{\mathrm{HF}}(R) 
    + \delta E_{\mathrm{int}}^{\mathrm{fc-CCSD}}(R) 
    + \delta E_{\mathrm{int}}^{\mathrm{fc-CCSDT}}(R) \\
    &+ \delta E_{\mathrm{int}}^{\mathrm{fc-CCSDTQ}}(R) 
    + \delta E_{\mathrm{int}}^{\mathrm{fc-(P)}}(R) 
    + \delta E_{\mathrm{int}}^{\mathrm{fc-FCI}}(R) \\
    &+ \delta E_{\mathrm{int}}^{\mathrm{ae-CCSDT}}(R)
    + \delta E_{\mathrm{int}}^{\mathrm{rel1}}(R) 
    + \delta E_{\mathrm{int}}^{\mathrm{rel2}}(R) \\
    &+ \delta E_{\mathrm{int}}^{\mathrm{QED}}(R) 
    + \delta E_{\mathrm{int}}^{\mathrm{DBOC}}(R).
\end{split}
\end{align}
Note that the definitions of each component given below refer to the best estimates of the complete basis set (CBS) limit of each quantity (basis set convergence and the corresponding extrapolations are discussed further in the text). The $E_{\mathrm{int}}^{\mathrm{HF}}(R)$ term is the contribution to the interaction energy calculated using the Hartree--Fock (HF) method. \cite{Hartree_1928,fock1930naherungsmethode} We employ the restricted Hartree--Fock formalism \cite{roothaan1951new} for the dimer and the restricted open-shell Hartree--Fock variant \cite{roothaan1960self} for the single atom. The $\delta E_{\mathrm{int}}^{\mathrm{fc-CCSD}}(R)$ term is the pure correlation contribution to the interaction energy calculated using the frozen-core (ten active valence electrons) CCSD method. Next, $\delta E_{\mathrm{int}}^{\mathrm{fc-CCSDT}}(R)$ is the difference in the interaction energy calculated with the frozen-core CCSDT and CCSD methods. Similarly, $\delta E_{\mathrm{int}}^{\mathrm{fc-CCSDTQ}}(R)$ denotes the difference between the frozen-core CCSDTQ \cite{kucharski1991recursive,oliphant1991coupled,kucharski1992coupled} and CCSDT methods, and $\delta E_{\mathrm{int}}^{\mathrm{fc-(P)}}(R)$ is the difference between CCSDTQ(P) \cite{kallay2001higher} and CCSDTQ. The final valence-only term is $\delta E_{\mathrm{int}}^{\mathrm{fc-FCI}}(R)$ which accounts for the difference between the frozen-core FCI and CCSDTQ(P) methods. This completes the list of corrections that account for the dominant valence-only correlated contribution to the interaction energy.

The next term, $\delta E_{\mathrm{int}}^{\mathrm{ae-CCSDT}}(R)$, takes into account the core-core and mixed core-valence correlation effects originating from the two $1s^2$ nitrogen core orbitals. This correction is defined as the difference between the all-electron and frozen-core CCSDT contributions to the interaction energy. The following three terms account for relativistic and QED corrections \cite{BeSal,Pachucki2004} of the order of $1/c^2$ and $1/c^3$, respectively, where $c$ is the speed of light. A detailed protocol used for the determination of these corrections is given further in the text. Note that the last two corrections, $\delta E_{\mathrm{int}}^{\mathrm{rel2}}(R)$ and $\delta E_{\mathrm{int}}^{\mathrm{QED}}(R)$, were neglected in the calculations reported in Paper~I based on the experience from earlier calculations for light atoms available in the literature. However, less is known about the importance of these terms in molecular systems, and the conclusions drawn from the available results are less definitive. \cite{Cencek2012,Lesiuk2019be} Therefore, it is worthwhile to take the terms $\delta E_{\mathrm{int}}^{\mathrm{rel2}}(R)$ and $\delta E_{\mathrm{int}}^{\mathrm{QED}}(R)$ into consideration, at least approximately, which would also provide valuable experience for other systems.

Finally, we note that all contributions discussed thus far were obtained within the Born--Oppenheimer (BO) approximation. \cite{born1927quantentheorie,kutzelnigg1997adiabatic} In this approximation all corrections of the order of electron-to-nucleus mass ratio are neglected. While this ratio is tiny for systems such as N$_2$, the influence of the post-BO effects can be taken into account by adding the so-called diagonal BO correction (DBOC), known also as the adiabatic correction. \cite{gauss2006analytic} This is the purpose of the last contribution included in Eq.~(\ref{composite}), namely $\delta E_{\mathrm{int}}^{\mathrm{DBOC}}(R)$.

The short-range part of the potential was calculated using the above protocol for a grid of internuclear distances centered approximately at the minimum of the potential energy curve located at $R_e\approx 2.07$. The starting point of the grid corresponds to about $0.7R_e$ which is already deep within the repulsive wall of the potential and inclusion of shorter interatomic distances is not necessary from the point of view of the spectroscopic parameters determined in this work. Indeed, for the smallest internuclear distance considered here, $R=1.45$, the potential lies about $10\,$eV above the dissociation limit or about $20\,$eV above the minimum of the potential. The last grid point corresponds to about $1.5 R_e$ which delimits the region where the single-reference CC method are expected to perform well. The whole grid consists of $37$ point and is the most dense near $R_e$ and gets progressively more spaced out as one moves away from $R_e$ in both directions. The explicit list of grid points is given in the Supplementary Material. \cite{supp} Further in the text, we will list explicit results only for three internuclear distances: $R=2.070$ (very close to the equilibrium distance) and two points, $R=1.450$ and $R=2.700$, which are roughly equidistant from the former in both directions. These results serve the illustrative purposes for the sake of the accompanying discussion. However, the complete set of results for all grid points are always given in the Supplementary Material. \cite{supp}

\subsection{Hartree--Fock contribution}
\label{sec:hf}

The Hartee--Fock contribution to the interaction energy was calculated using the aug-cc-pV$X$Z basis set family taken from the literature. \cite{dunning89,thorpe21} The same basis sets were employed for calculation of the valence-only contributions reported in the next sections. Basis sets of quality ranging from double- ($X=2$) to octuple-zeta ($X=8$) were optimized for nitrogen in Refs.~\onlinecite{dunning89,thorpe21}. Further in the text we refer to the quantity $X$ as the cardinal number. Note that we use only the singly-augmented variants (aug-) of these basis sets here. Based on preliminary calculations we found that double augmentation of these basis sets, similar to the one used in Paper~I, does not lead to appreciable increase in the accuracy in calculation of the short-range part of the potential. Moreover, increase of the augmentation level results in significant linear dependencies within the basis set and should be avoided to preserve the stability of the results. Similar conclusions hold when the basis sets are supplemented by a set of midbond functions (located at the center of symmetry of the molecule) which we also tested at a preliminary stage. Psi4 program has been used to calculate the Hartree--Fock contributions. \cite{smith2020psi4}

In Table~\ref{tab:hf}, we provide Hartee-Fock contribution to the interaction energy for selected internuclear distances calculated with the aug-cc-pV$X$Z basis sets with $X=6,7,8$. It is known that $E_{\mathrm{int}}^{\mathrm{HF}}(R)$ usually converges fast with respect to the basis set size and is sometimes extrapolated to the CBS limit using three-point exponential formulas. However, in our case we find that such extrapolation is not necessary. The difference between $E_{\mathrm{int}}^{\mathrm{HF}}(R)$ evaluated with the two largest basis sets never exceeds $1\,$~meV which is a small error from the point of view of this work. However, as the $E_{\mathrm{int}}^{\mathrm{HF}}(R)$ contribution is particularly large, we provide an independent verification that this fast convergence is not an artifact of some systematic basis set error. To this end, we double-check these data using numerical Hartee-Fock results obtained with grid-based methods which do not rely on the conventional basis set expansions. 
For the diatom, we employed the {\sc x2dhf} program by Kobus \cite{kobus13} which has recently been revised \cite{kobus25} and is publicly available. \cite{x2dhf} It relies on a finite-difference approach defined in a transformed ellipsoidal coordinate system with variables $\nu$ and $\mu$, see Ref.~\onlinecite{kobus25} for technical details. In our calculations we employed several numerical grids for $\nu$, ranging from a coarse grid of 170 points (with a practical value of infinity equal to 40 bohr) to a finer grid of nearly 400 points (where practical infinity was set to 100 bohr). During the calculations the number of grid points for the variable $\mu$ were automatically determining by the {\sc x2dhf} program based on previously mentioned values. \cite{x2dhf} The convergence with respect to the grid was conducted for each internuclear distance and the final values were converged to within several nanohartrees. The Hartree--Fock energy of the nitrogen atom was taken from the work of Saito \cite{saito09} which was calculated using the grid-based $B$-splines expansion \cite{saito03} with error smaller than one nanohartree.  Therefore, the ``numerical'' interaction energies reported in Table~\ref{tab:hf} are certain to all digits shown. We see a good agreement with the results obtained with the aug-cc-pV8Z basis set and the discrepancies do not exceed $1\,$meV. As the ``numerical'' Hartree--Fock results are slightly more accurate, we chose them as our recommended value of Hartree--Fock interaction energies. We assign a conservative uncertainty of $10^{-4}$~eV to this term for all internuclear distances.

\begin{table}
\caption{\label{tab:hf}
Interaction energies of N$_2$ calculated using the restricted Hartree--Fock method within the aug-cc-pV$X$Z basis set family ($X=6,7,8$) as well as numerical grid-based methods (last row) for internuclear distances $R=1.450$, $R=2.070$, and $R=2.700$. The interaction energies are given in eV and the internuclear distances in bohrs.
}
\begin{ruledtabular}
\begin{tabular}{cccc}
 $X$ & $R=1.450$ & $R=2.070$ & $R=2.700$ \\
 \hline\\[-1.2em]
6 & $-$12.6634 & 5.217\,48 & $-$0.833\,016 \\
7 & $-$12.6620 & 5.217\,80 & $-$0.832\,962 \\
8 & $-$12.6615 & 5.217\,92 & $-$0.832\,911 \\
\hline\\[-1.2em]
numerical & $-$12.6611 & 5.217\,99 & $-$0.832\,879 \\
\end{tabular}
\end{ruledtabular}
\end{table}

\subsection{Frozen-core CCSD contribution}
\label{sec:fc-ccsd}

Using the Psi4 program, we calculated the frozen-core CCSD contribution to the interaction energy, $\delta E_{\mathrm{int}}^{\mathrm{fc-CCSD}}(R)$, within the aug-cc-pV$X$Z basis set family with $X=6,7,8$. The results are reported in Table~\ref{tab:fcccsd} (together with their respective uncertainties) for three selected internuclear distances. The corresponding data for the remaining grid points are included in the Supplementary Material. \cite{supp} The uncertainties are substantial (ca. 0.3\%) for the grid points within the repulsive wall of the potential, but this is acceptable as the points in this range do not have a major influence on the spectroscopic parameters determined later. As the internuclear distance is increased, the uncertainties drop sharply reaching the level of roughly 0.05\% near the bottom of the potential and further decrease slowly with $R$. 

\begin{table}
\caption{\label{tab:fcccsd}
Frozen-core CCSD contribution to the interaction energy of N$_2$, $\delta E_{\mathrm{int}}^{\mathrm{fc-CCSD}}(R)$, calculated within the aug-cc-pV$X$Z basis set family for internuclear distances $R=1.450$, $R=2.070$, and $R=2.700$. The extrapolated results and their expanded $(k=2)$ uncertainty estimates are given in the last row. The interaction energies are given in eV and the internuclear distances in bohrs.
}
\begin{ruledtabular}
\begin{tabular}{cccc}
 $X$ & $R=1.450$ & $R=2.070$ & $R=2.700$ \\
 \hline\\[-1.2em]
6 & 3.0497 & 4.1873 & 6.0572 \\
7 & 3.0800 & 4.2041 & 6.0676 \\
8 & 3.0955 & 4.2134 & 6.0735 \\
\hline\\[-1.2em]
$\infty$ & 3.1297(109) & 4.2338(28) & 6.0866(9) \\
\end{tabular}
\end{ruledtabular}
\end{table}

\subsection{Frozen-core CCSDT contribution}
\label{sec:fc-ccsdt}

The next major contribution to the interaction energy, $\delta E_{\mathrm{int}}^{\mathrm{fc-CCSDT}}(R)$, takes into account triple excitations with respect to the reference determinant. It was calculated using the CCSDT method within CFOUR program \cite{matthews2020coupled,cfour} using the same aug-cc-pV$X$Z basis set family as in the preceding sections, but the largest basis set we could employ for $\delta E_{\mathrm{int}}^{\mathrm{fc-CCSDT}}(R)$ is $X=6$ due to the increasing computational costs and memory requirements of the CCSDT method. In Table~\ref{tab:fcccsdt} we report the calculated $\delta E_{\mathrm{int}}^{\mathrm{fc-CCSDT}}(R)$ contributions for three selected internuclear distances along with their uncertainty estimates. The estimated absolute uncertainties are remarkably stable as a function of $R$ while relative errors range from few percent for small internuclear distances to 0.5\% and less for larger $R$. In comparison with the CCSD contribution, the $\delta E_{\mathrm{int}}^{\mathrm{fc-CCSDT}}(R)$ term is at least by an order of magnitude smaller and despite the larger relative uncertainties of the latter, both terms have similar absolute uncertainties for small and medium $R$. For larger values of $R$, the uncertainty of $\delta E_{\mathrm{int}}^{\mathrm{fc-CCSDT}}(R)$ dominates over $\delta E_{\mathrm{int}}^{\mathrm{fc-CCSD}}(R)$.

\begin{table}
\caption{\label{tab:fcccsdt}
Frozen-core CCSDT contribution to the interaction energy of N$_2$, $\delta E_{\mathrm{int}}^{\mathrm{fc-CCSDT}}(R)$, calculated within the aug-cc-pV$X$Z basis set family for internuclear distances $R=1.450$, $R=2.070$, and $R=2.700$. The extrapolated results and their expanded $(k=2)$ uncertainty estimates are given in the last row. The interaction energies are given in eV and the internuclear distances in bohrs.
}
\begin{ruledtabular}
\begin{tabular}{cccc}
 $X$ & $R=1.450$ & $R=2.070$ & $R=2.700$ \\
 \hline\\[-1.2em]
4 & 0.1260 & 0.3611 & 0.7753 \\
5 & 0.1322 & 0.3672 & 0.7827 \\
6 & 0.1342 & 0.3692 & 0.7851 \\
\hline\\[-1.2em]
$\infty$ & 0.1372(39) & 0.3723(36) & 0.7888(44) \\
\end{tabular}
\end{ruledtabular}
\end{table}

One may argue that to improve accuracy of the triple-excitation contribution to the excitation energy it is possible to employ the CCSD(T) method \cite{RAGHAVACHARI1989479,BARTLETT1990513,STANTON1997130} as an intermediate step. It is known that this method is highly successful in reproduction of the triple-excitation effects at a significantly lower cost. For example, the total T contribution can be split into two terms: (i) difference between CCSD(T) and CCSD methods and (ii) difference between CCSDT and CCSD(T). The second contribution is expected to be smaller than the first while the first can be calculated in a larger basis set due to the decreased cost of the CCSD(T) method. Unfortunately, we found that this approach is successful only for small values of $R$, where the uncertainty of $\delta E_{\mathrm{int}}^{\mathrm{fc-CCSDT}}(R)$ is not dominant anyway. As we increase the internuclear distance, the quality of CCSD(T) results quickly deteriorate, likely due to the quasi-perturbative nature of this method. Therefore, we bypassed this possibility, and, to avoid additional technical complications, the total $\delta E_{\mathrm{int}}^{\mathrm{fc-CCSDT}}(R)$ correction is calculated using a one-step procedure.

\subsection{Frozen-core CCSDTQ contribution}
\label{sec:fc-ccsdtq}

Next, we move on to the $\delta E_{\mathrm{int}}^{\mathrm{fc-CCSDTQ}}(R)$ term which accounts for the quadruple-excitation contributions. In determination of this term we are limited to the $X=4$ basis set, but the remaining details of the calculations and uncertainty estimation are the same as for the preceding correlation contributions. The results were obtained using MRCC program \cite{mester2025overview,mrcc} interfaced to CFOUR and are shown in Table~\ref{tab:fcccsdtq} for the selected three nuclear distances. On average, the $\delta E_{\mathrm{int}}^{\mathrm{fc-CCSDTQ}}(R)$ term is by an order of magnitude smaller than $\delta E_{\mathrm{int}}^{\mathrm{fc-CCSDT}}(R)$ and similarly grows as the internuclear distance is increased. However, due to the smaller $X=2,3,4$ basis sets available, the uncertainties of these two terms are similar.

\begin{table}
\caption{\label{tab:fcccsdtq}
Frozen-core CCSDTQ contribution to the interaction energy of N$_2$, $\delta E_{\mathrm{int}}^{\mathrm{fc-CCSDTQ}}(R)$, calculated within the aug-cc-pV$X$Z basis set family for internuclear distances $R=1.450$, $R=2.070$, and $R=2.700$. The extrapolated results and their expanded $(k=2)$ uncertainty estimates are given in the last row. The interaction energies are given in eV and the internuclear distances in bohrs.
}
\begin{ruledtabular}
\begin{tabular}{cccc}
 $X$ & $R=1.450$ & $R=2.070$ & $R=2.700$ \\
 \hline\\[-1.2em]
2 & 0.0072 & 0.0408 & 0.1516 \\
3 & 0.0072 & 0.0411 & 0.1614 \\
4 & 0.0081 & 0.0425 & 0.1633 \\
\hline\\[-1.2em]
$\infty$ & 0.0090(27) & 0.0438(36) & 0.1650(35) \\
\end{tabular}
\end{ruledtabular}
\end{table}

\subsection{Frozen-core CCSDTQ(P) and FCI contributions}
\label{sec:fc-fci}

The last two valence-only corrections to the interaction energy, $\delta E_{\mathrm{int}}^{\mathrm{fc-(P)}}(R)$ and $\delta E_{\mathrm{int}}^{\mathrm{fc-FCI}}(R)$, take into account excitations higher than quadruple and were obtained through MRCC program interfaced to CFOUR, and FCI code developed by one of the authors (M.P.). \cite{hector} By observing that contributions to $E_{\mathrm{int}}(R)$ decrease roughly by an order of magnitude as the excitation level is increased (D$\rightarrow$T, T$\rightarrow$Q, etc.) we expect these terms to be small, but they still may be non-negligible with the present accuracy requirements.

The first component, $\delta E_{\mathrm{int}}^{\mathrm{fc-(P)}}(R)$, is calculated using the CCSDTQ(P) method and approximately accounts for the effects of pentuple excitations which dominate the total post-CCSDTQ correction, especially for smaller $R$. The second component, $\delta E_{\mathrm{int}}^{\mathrm{fc-FCI}}(R)$, is the difference between CCSDTQ(P) and FCI results within the same basis set. The reason for this separation is the fact that the CCSDTQ(P) contribution can be calculated with a larger basis set. Indeed, despite the large cost, we managed to obtain the CCSDTQ(P) results with $X=2,3$ basis sets. The final values of $\delta E_{\mathrm{int}}^{\mathrm{fc-(P)}}(R)$ are obtained using the Riemann extrapolation from this pair of basis sets. Unfortunately, as only two basis sets are available, we cannot estimate the uncertainty using the random walk procedure that was used in the preceding sections (for that, we would need data from three consecutive $X$). Therefore, the uncertainty of the CCSDTQ(P) component was estimated more conservatively as a difference between the extrapolated result and the corresponding data obtained with $X=3$ basis set. The results are reported in Table~\ref{tab:postq} for the selected set of internuclear distances. As a technical note, we add that the contribution of the $(P)$ correction to the total energy of an isolated nitrogen atom is minuscule (in the frozen-core approximation). To avoid performing computationally expensive open-shell CCSDTQ(P) calculations for each internuclear separation, we simply assumed that the atomic $(P)$ correction is equal to zero in the final determination of the $\delta E_{\mathrm{int}}^{\mathrm{fc-(P)}}(R)$ contribution to the interaction energy.

\begin{table}
\caption{\label{tab:postq}
Two post-CCSDTQ corrections to the interaction energy of N$_2$, $\delta E_{\mathrm{int}}^{\mathrm{fc-(P)}}(R)$ and $\delta E_{\mathrm{int}}^{\mathrm{fc-FCI}}(R)$, for internuclear distances $R=1.450$, $R=2.070$, and $R=2.700$. See the main text for details of the extrapolation and uncertainty estimation. The interaction energies are given in eV and the internuclear distances in bohrs.
}
\begin{ruledtabular}
\begin{tabular}{cccc}
 $X$ & $R=1.450$ & $R=2.070$ & $R=2.700$ \\
 \hline\\[-1.2em]
 \multicolumn{4}{c}{$\delta E_{\mathrm{int}}^{\mathrm{fc-(P)}}(R)$} \\
 \hline\\[-1.2em]
 2 & 0.00025 & 0.0056 & 0.0458 \\
 3 & 0.00015 & 0.0046 & 0.0442 \\
 \hline\\[-1.2em]
 $\infty$ & 0.00009(6) & 0.0040(6) & 0.0433(10) \\
 \hline\\[-1.2em]
 \multicolumn{4}{c}{$\delta E_{\mathrm{int}}^{\mathrm{fc-FCI}}(R)$} \\
 \hline
 2(mod) & 0.00019 & 0.00017 & $-$0.0069 \\
 \hline\\[-1.2em]
 $\infty$ & 0.000071(36) & 0.00012(6) & $-$0.0067(34) \\
\end{tabular}
\end{ruledtabular}
\end{table}

Moving on to the $\delta E_{\mathrm{int}}^{\mathrm{fc-FCI}}(R)$ contribution, calculation of this quantity using the FCI method is extremely computationally expensive. We did not manage to evaluate this correction even in the smallest ($X=2$) basis used in this work. Therefore, we generated a truncated basis set which is a slight modification of the parent aug-cc-pV2Z as follows. The standard aug-cc-pV2Z basis includes two $d$-type functions with exponents $0.8170$ and $0.2300$. Taking into consideration that each $d$ function has five spherical components, elimination of one of them would significantly reduce the size of the basis set (from 46 to 36 functions for the whole molecule) and make the FCI computations feasible, albeit at a huge cost. We initially considered just dropping the $d$ function that has a smaller contribution to the interaction energy at the CCSDTQ or CCSDTQ(P) level. However, we found that a better approach is to replace the two original $d$ functions by a single $d$ function with exponent equal to the geometric average of the original exponents. This leads to a truncated basis, which we denote as aug-cc-pV2Z(mod) or 2(mod) for short, with a single $d$ function with the exponent equal to $0.4335$. 

We found that the truncated basis performs well in reproduction of higher-order correlation effects. For example, the $\delta E_{\mathrm{int}}^{\mathrm{fc-(P)}}(R)$ contribution for $R=2.070$ is $5.65$~meV and $5.73$~meV within the aug-cc-pV2Z and aug-cc-pV2Z(mod) basis set, respectively, a difference of a mere $1-2\%$. This gives us confidence that the latter basis is just as capable as the former in reproducing higher-order correlation contributions to the interaction energy. Consequently, we proceeded to evaluate the $\delta E_{\mathrm{int}}^{\mathrm{fc-FCI}}(R)$ correction within the aug-cc-pV2Z(mod) basis. However, having only data from a single basis set prevents us from performing extrapolation to the CBS limit and hence estimating the uncertainty in a way similar as for $\delta E_{\mathrm{int}}^{\mathrm{fc-(P)}}(R)$. To obtain an estimate of the CBS limit, we assume that the $\delta E_{\mathrm{int}}^{\mathrm{fc-(P)}}(R)$ and $\delta E_{\mathrm{int}}^{\mathrm{fc-FCI}}(R)$ converge with the basis set size at the same rate. With this assumption, the CBS limit of the latter term can be obtained by scaling the $\delta E_{\mathrm{int}}^{\mathrm{fc-FCI}}(R)$ results calculated within the aug-cc-pV2Z(mod) basis by the ratio of the $\delta E_{\mathrm{int}}^{\mathrm{fc-(P)}}(R)$ term obtained in the CBS limit and in the aug-cc-pV2Z(mod) basis. However, as we are not able to independently verify the reliability of this \emph{ad hoc} procedure, we assign a very conservative 50\% relative uncertainty level to the $\delta E_{\mathrm{int}}^{\mathrm{fc-FCI}}(R)$ contribution obtained in this way. The final values of the FCI correction for selected internuclear distances are given in Table~\ref{tab:postq}. We see that this term is completely negligible near the minimum of the potential energy curve and within the repulsive wall. However, it grows substantially with $R$ and becomes substantial as we move to larger distances.

\subsection{Core-core and core-valence effects}
\label{sec:ae-ccsdt}

It is known that in most molecules valence electrons are responsible for the vast majority of the correlation contribution to the interaction energy. However, with the present accuracy requirements the correlation between core electrons and mixed core-valence contributions cannot be safely neglected. To establish a proper protocol for determination of these contributions, we note that the CCSDT method recovers more than 95\% of the valence correlation effects for all internuclear distances considered here. As we will see shortly, the core-core and core-valence effects are by two to three orders of magnitude smaller than the pure valence contribution. As a result, the CCSDT method is a sufficient level of theory for determination of the former contributions and post-CCSDT effects are negligible in comparison with other sources of error.

In Table~\ref{tab:aeccsdt} we report the results of the calculations of the core-core and core-valence contributions to the interaction energy obtained using CFOUR program. In these calculations we employ the aug-cc-pCV$X$Z basis set family, $X=3,4,5$, taken from Ref.~\onlinecite{woon1995gaussian}. In comparison with the valence-only aug-cc-pV$X$Z basis sets, they incorporate an additional set of ``tight'' functions with large exponents that improve the description of the regions close to the atomic nuclei. From Table~\ref{tab:aeccsdt} we see that the core correlation effects are substantial for smaller internuclear distances, but decay rapidly as the atoms are further apart. However, the $\delta E_{\mathrm{int}}^{\mathrm{ae-CCSDT}}(R)$ correction is non-negligible within the present accuracy target and has to be included for a balanced description of the total interaction energy. Interestingly, unlike the valence-only contributions considered above, the $\delta E_{\mathrm{int}}^{\mathrm{ae-CCSDT}}(R)$ term changes sign as a function of $R$. It is attractive for small internuclear distances, but crosses zero around $R=2.3-2.4$ and becomes repulsive for larger $R$.

\begin{table}
\caption{\label{tab:aeccsdt}
All-electron CCSDT contribution to the interaction energy of N$_2$, $\delta E_{\mathrm{int}}^{\mathrm{ae-CCSDT}}(R)$, calculated within the aug-cc-pCV$X$Z basis set family for internuclear distances $R=1.450$, $R=2.070$, and $R=2.700$. The extrapolated results and their expanded $(k=2)$ uncertainty estimates are given in the last row. The interaction energies are given in eV and the internuclear distances in bohrs.
}
\begin{ruledtabular}
\begin{tabular}{cccc}
 $X$ & $R=1.450$ & $R=2.070$ & $R=2.700$ \\
 \hline\\[-1.2em]
3 & 0.1516 & 0.0269 & $-$0.0147 \\
4 & 0.1976 & 0.0332 & $-$0.0181 \\
5 & 0.2158 & 0.0359 & $-$0.0191 \\
\hline\\[-1.2em]
$\infty$ & 0.2382(22) & 0.0393(6) & $-$0.0202(16) \\
\end{tabular}
\end{ruledtabular}
\end{table}

Finally, as anticipated above, inclusion of higher excitations is not necessary in determination of the core correlation effects. For small internuclear distances, where $\delta E_{\mathrm{int}}^{\mathrm{ae-CCSDT}}(R)$ is relatively large, they are not expected to exceed more than 1\% of the total core contribution based on the corresponding ratio in the valence-only results. Therefore, they are almost certainly smaller than 0.01\% of the total interaction energy when all contributions are summed over. For larger internuclear distances, the post-CCSDT effects may reach a few percent of $\delta E_{\mathrm{int}}^{\mathrm{ae-CCSDT}}(R)$, but for large $R$ the total core correlation contribution is tiny (a few hundredths of eV) and hence these effects would drown in the uncertainty of $\delta E_{\mathrm{int}}^{\mathrm{ae-CCSDT}}(R)$.

\subsection{Relativistic and QED corrections}
\label{sec:rel12}

Calculation of the relativistic corrections to the potential energy curve of N$_2$ in this work is based on the perturbative approach using the Breit--Pauli (BP) Hamiltonian. \cite{BeSal,Pachucki2004} Similarly as discussed in Paper~I, expectation values of all spin-dependent terms in the BP Hamiltonian either vanish or can be rewritten as a combination of spin-free corrections, because we consider the ground $^1\Sigma_g^+$ state of the diatom and $^4S$ state of the free atom. We are therefore left with two one-electron spin-independent relativistic corrections:
\begin{align}
\label{rel1}
    E_{\mathrm{rel1}}(R) = 
    -\frac{1}{8c^2}\,\langle \sum_i\nabla_i^4 \rangle
    +\frac{\pi}{2c^2}\,\langle \sum_{iA} Z_A\,\delta(\mathbf{r}_{iA}) \rangle,
\end{align}
where $\delta(\mathbf{r})$ is the Dirac delta distribution, $\nabla_i$ is the gradient operator acting on the electronic coordinates, while the symbol $\langle X\rangle$ denotes the expectation value of an operator $X$ on the ground-state non-relativistic wavefunction in the BO approximation. In the above equation we have introduced the following notation: the indices $i$ and $A$ denote the electrons and nuclei in the system, respectively, $\mathbf{r}_i$ are the coordinates of the $i$-th electron, $\mathbf{r}_A$ are the coordinates of the $A$-th nucleus with charge $Z_A$, and $\mathbf{r}_{iA} = \mathbf{r}_i - \mathbf{r}_A$. The two terms defined in Eq.~(\ref{rel1}) are usually referred to as the mass-velocity and the one-electron Darwin corrections in the literature.

There are also two spin-free two-electron relativistic corrections originating from the BP Hamiltonian:
\begin{align}
\label{rel2}
\begin{split}
    E_{\mathrm{rel2}}(R) &= 
    \frac{\pi}{c^2}\,\langle \sum_{i>j} \delta(\mathbf{r}_{ij}) \rangle \\
    &+ \frac{1}{2c^2}\,\langle \sum_{i>j} \left[ \frac{\nabla_i\cdot\nabla_j}{r_{ij}} +
    \frac{\mathbf{r}_{ij}(\mathbf{r}_{ij}\cdot\nabla_i)\cdot\nabla_j}{r_{ij}^3}\right] \rangle,
\end{split}
\end{align}
which are usually called the two-electron Darwin and the orbit-orbit corrections, respectively. To avoid confusion, we point out that the first term defined above includes also the spin-spin term from the general Breit-Pauli Hamiltonian, hence the reverted sign in front of the two-electron Darwin correction. In these definitions we use the symbols $\mathbf{r}_{ij} = \mathbf{r}_{i} - \mathbf{r}_{j}$ and $r_{ij}=|\mathbf{r}_{ij}|$ for the inter-electronic distance. In the case of both one- and two-electron relativistic corrections, their contributions to the interaction energy, denoted by $\delta E_{\mathrm{int}}^{\mathrm{rel1}}(R)$ and $\delta E_{\mathrm{int}}^{\mathrm{rel2}}(R)$ in Eq.~(\ref{composite}), are calculated using the formula analogous to Eq.~(\ref{eint}) based on the expectation values of each operator evaluated for the diatom and the free atom separately.

The one-electron relativistic corrections were previously evaluated in Ref.~\onlinecite{lesiuk2023atomic} using internally contracted multireference configuration-interaction (ic-MRCI) method \cite{werner82,werner88,knowles88}. However, no uncertainty estimates were provided for them. To fill this gap, we used the raw data reported in Ref.~\onlinecite{lesiuk2023atomic} calculated with the uncontracted d-aug-cc-pV$X$Z, $X=4,5,6$, and estimated the uncertainty of the extrapolated results using the random walk procedure that is used in this work for all remaining quantities. In Table~\ref{tab:short} we report the final results together with their uncertainty estimates for the selected internuclear distances. We see that the one-electron relativistic corrections are not large (below $10\,$~meV for $R\ge1.700$), but are not negligible in comparison with other sources of error and have to be included for a balanced description of the potential energy curve. The uncertainty of $\delta E_{\mathrm{int}}^{\mathrm{rel1}}(R)$ is negligible in comparison with other terms, i.e., the data provided in Ref.~\onlinecite{lesiuk2023atomic} is of sufficient quality for the purposes of the present work.

In contrast with the one-electron relativistic corrections, we have not found any data in the literature for two-electron relativistic contributions. However, as we anticipate these corrections to be smaller than $\delta E_{\mathrm{int}}^{\mathrm{rel1}}(R)$, they do not have to be calculated with as high precision. To determine $\delta E_{\mathrm{int}}^{\mathrm{rel2}}(R)$ we used the multiconfigurational self-consistent field method (MCSCF) \cite{roos1980complete}, because the expectation values calculated using the coupled-cluster theory were significantly less stable than the corresponding energies. The the d-aug-cc-pV6Z basis set was used in these calculations and further extension of the basis changes values of the two-electron corrections at the third or fourth significant digit which is irrelevant in the present context. The two-electron Darwin and orbit-orbit integrals in the Gaussian basis necessary for these calculations were imported from the Dalton program package \cite{dalton}, while the MCSCF calculations were performed using the same code that was used for FCI calculations, see Sec.~\ref{sec:fc-fci}.

A more problematic question is the choice of the active space in the MCSCF method. In the subsequent discussion we will refer to the complete active space obtained by distribution of $n$ electrons into $m$ orbitals by the symbol $[n/m]$. We tested several physically-justified active spaces starting with the minimalistic valence-bond $[6/6]$, extending it further and checking how the result change with inclusion of additional orbitals. Some active spaces, despite no obvious deficiencies, suffered from lack of continuity of the results as a function of the internuclear distance. We finally settled on the active space $[10/22]$ which provides sufficiently converged results and no pathological behavior. This active space includes the full set of valence bonding ($\sigma$, $\pi$) and antibonding ($\sigma^*$, $\pi^*$) orbitals of N$_2$ plus: (i) all $\sigma$, $\sigma^*$, and $\pi$, $\pi^*$, orbitals originating from the combination of $3s$ and $3p$ atomic orbitals of nitrogen, (ii) all $\sigma$, $\sigma^*$, and $\delta$, $\delta^*$ molecular orbitals coming from the combination of $3d$ atomic orbitals. 

The final results for $\delta E_{\mathrm{int}}^{\mathrm{rel2}}(R)$ obtained at the MCSCF$[10/22]$ level of theory are shown in Table~\ref{tab:short}. In order to estimate the uncertainty of these results, we first note that the non-relativistic interaction energy calculated at the MCSCF$[10/22]$ agrees to better than 5\% with the best theoretical estimates for all internuclear distances under consideration. However, we acknowledge the fact that the error in calculation of the properties is frequently larger than for the energy and hence increase the relative uncertainty level by a factor of three (to 15\%). We see from Table~\ref{tab:short} that, somewhat surprisingly, the contribution of the two-electron relativistic effects to the interaction energy for some internuclear distances is only $2-3$ times smaller than of the one-electron relativistic corrections. Therefore, the relative importance of the two-electron effects appears to be larger for diatomic systems than for isolated atoms for which more data is available in the literature. From Table~\ref{tab:short} we also conclude that despite the uncertainty of $\delta E_{\mathrm{int}}^{\mathrm{rel2}}(R)$ was estimated rather conservatively, it does not contribute much to the overall error budget.

Finally, we consider the QED corrections to the interaction energy. It is known that the leading (of the order $1/c^3$ and $\ln c / c^3$) one-electron QED correction to the energy takes the form: \cite{eides2001theory}
\begin{align}
\label{qed}
    E_{\mathrm{QED}}(R) = \frac{4}{3c^3}\left( \frac{19}{30} + 2\ln c - \ln k_0 \right)
    \langle \sum_{iA} Z_A\,\delta(\mathbf{r}_{iA}) \rangle,
\end{align}
where $\ln k_0$ is the so-called Bethe logarithm. \cite{BeSal} We neglected here the two-electron QED corrections including the so-called Araki--Sucher term. \cite{araki57,sucher58} Methods for calculation of these quantities have been reported in the literature, \cite{balcerzak17} but they were estimated to be by an order of magnitude smaller than the one-electron QED contribution and hence negligible here. In evaluation of the correction given by Eq.~(\ref{qed}) we adopt an additional approximation to the $\ln k_0$ term, exploiting the fact that this quantity is weakly dependent on the internuclear distance (at least for the range of $R$ considered here). Therefore, we replace the molecular $\ln k_0$ by the value corresponding to the $R\rightarrow\infty$ limit, i.e., the Bethe logarithm for a single nitrogen atom. We use the atomic value of $\ln k_0 = 6.973$ reported in Ref.~\onlinecite{lesiuk2023atomic}. Note that the expectation value of the one-electron Dirac delta distribution present in Eq.~(\ref{qed}) has already been evaluated above when the one-electron relativistic effects were considered. Therefore, within the approximate scheme outlined above, inclusion of the dominant QED correction requires no new quantities to be determined. To take into account all of the approximations involved in evaluation of Eq.~(\ref{qed}), we assign a conservative 50\% uncertainty estimate to this contribution to the interaction energy of N$_2$. We final results for the selected set of internuclear distances are reported in Table~\ref{tab:short}. We see that the QED contributions to the interaction energy are tiny (below $1\,$~meV for all internuclear distances considered) and can be neglected within the present accuracy requirements.

\subsection{Diagonal Born--Oppenheimer correction}
\label{sec:dboc}

All calculations reported thus far in this work have been performed within the clamped-nuclei BO framework.  However, it is known that this approximation neglects post-BO effects that depend on the inverse of the electron-to-nucleus mass ratio. The leading post-BO contribution is the so-called diagonal BO correction (DBOC) given by the formula: \cite{Handy1986,Handy1996}
\begin{align}
    E_{\mathrm{DBOC}}(R) = \langle -\sum_A \frac{1}{2M_A}\nabla_A^2 \rangle,
\end{align}
where $\nabla_A$ is the gradient operator acting on the coordinates of the $A$-th nucleus (with mass $M_A$), and the expectation value is evaluated with the non-relativistic BO wavefunction. Here and throughout this work, we consider only the $^{14}$N isotope with the highest natural abundance ($>99\%$) and atomic mass $M(^{14}\mathrm{N}) = 14.003\,074$ in the atomic mass units. \cite{Huang2021,Wang2021} Note that we use the atomic mass in evaluation of DBOC, as well as for subsequent calculations of rovibrational levels, rather than the ``bare'' nuclear mass. This approach is known to effectively take into account also higher-order post-BO corrections and provides a better description of the total post-BO effects. \cite{Przybytek2017}

To estimate the magnitude of the DBOC correction we evaluated it using the formalism developed in Ref.~\onlinecite{gauss2006analytic} at the CCSD/aug-cc-pCVQZ level of theory with all electrons correlated using CFOUR program package \cite{matthews2020coupled,cfour}. Due to the limitations of the implementation, unrestricted Hartree--Fock reference was used in the calculation of DBOC, but the effect of spin-contamination are small for the internuclear distances considered and hence the impact of this choice is expected to be marginal. To take into account the incompleteness of the basis set and the possible post-CCSD contributions, we assign a large 50\% relative uncertainty level to the calculated values. The results are included in Table~\ref{tab:short} for the selected internuclear distances. One can see that near the minimum of the potential, the DBOC contributes no more than $1\,$~meV to the interaction energy and is even smaller than the QED contribution. Only for very small internuclear distances, the $\delta E_{\mathrm{int}}^{\mathrm{DBOC}}(R)$ contribution slightly exceeds $1\,$~meV, but is still negligible in comparison with the uncertainties of other terms. Therefore, for consistency with previous considerations, we will neglect the DBOC contribution in the construction of the final potential energy curve.

\subsection{The final results and error budget}
\label{sec:sum-short}

\begin{table}
\caption{\label{tab:short}
Summary of the results for the short-range part of the potential for internuclear distances $R=1.450$, $R=2.070$, and $R=2.700$. The interaction energies are given in eV and the internuclear distances in bohrs. Expanded $(k=2)$ uncertainty estimates of each term and the sum are given in parentheses.
}
\begin{ruledtabular}
\begin{tabular}{llll}
 contribution & \multicolumn{1}{c}{$R=1.450$} & \multicolumn{1}{c}{$R=2.070$} & \multicolumn{1}{c}{$R=2.700$} \\
 \hline\\[-1.2em]
 $E_{\mathrm{int}}^{\mathrm{HF}}(R)$               
 & $-$12.6611(1) & \phantom{$-$}5.2180(1) & $-$0.8329(1) \\
 $\delta E_{\mathrm{int}}^{\mathrm{fc-CCSD}}(R)$   
 & \phantom{$-$0}3.1297(109) & \phantom{$-$}4.2338(28) & \phantom{$-$}6.0866(9) \\
 $\delta E_{\mathrm{int}}^{\mathrm{fc-CCSDT}}(R)$  
 & \phantom{$-$0}0.1372(39) & \phantom{$-$}0.3723(36) & \phantom{$-$}0.7888(44) \\
 $\delta E_{\mathrm{int}}^{\mathrm{fc-CCSDTQ}}(R)$ 
 & \phantom{$-$0}0.0090(27) & \phantom{$-$}0.0438(36) & \phantom{$-$}0.1650(35) \\
 $\delta E_{\mathrm{int}}^{\mathrm{fc-(P)}}(R)$    
 & \phantom{$-$0}0.00009(6) & \phantom{$-$}0.0040(6) & \phantom{$-$}0.0433(10) \\
 $\delta E_{\mathrm{int}}^{\mathrm{fc-FCI}}(R)$    
 & \phantom{$-$0}0.000071(36) & \phantom{$-$}0.00012(6) & $-$0.0067(34) \\
 $\delta E_{\mathrm{int}}^{\mathrm{ae-CCSDT}}(R)$  
 & \phantom{$-$0}0.2382(22) & \phantom{$-$}0.0393(6) & $-$0.0202(16) \\
 $\delta E_{\mathrm{int}}^{\mathrm{rel1}}(R)$      
 & \phantom{0$-$}0.0311(2) & $-$0.0063(1) & $-$0.0057(1) \\
 $\delta E_{\mathrm{int}}^{\mathrm{rel2}}(R)$      
 & \phantom{0}$-$0.0137(21) & $-$0.0021(3) & \phantom{$-$}0.00084(13) \\
 $\delta E_{\mathrm{int}}^{\mathrm{QED}}(R)$       
 & \phantom{0}$-$0.00016(8) & \phantom{$-$}0.00042(21) & \phantom{$-$}0.00012(6)\phantom{0} \\
 $\delta E_{\mathrm{int}}^{\mathrm{DBOC}}(R)$      
 & \phantom{0}$-$0.0018(9) & \phantom{$-$}0.00029(15) & \phantom{$-$}0.00029(15) \\
\hline\\[-1.2em]
total $E_{\mathrm{int}}(R)$ 
& \phantom{0}$-$9.1294(123) & \phantom{$-$}9.9029(59) & \phantom{$-$}6.2191(69) \\
\end{tabular}
\end{ruledtabular}
\end{table}

In Table~\ref{tab:short} we provide the summary of the results obtained for the short-range part of the potential for three representative internuclear distances. This enables us to judge how the estimated uncertainties of the individual terms of the composite scheme defined in Eq.~(\ref{composite}) impact the overall uncertainty. Assuming no statistical correlation, the total uncertainties are evaluated using the conventional error propagation rules, i.e., by summing the squares of uncertainties in each term and taking the square root of the sum. We reiterate that the expanded ($k=2$) uncertainties are reported here.

We see that in the repulsive part of the potential ($R=1.450$) the uncertainty of the frozen-core CCSD contribution is the dominant. Therefore, potential improvements are possible here in the future by using a larger basis set for the determination of the $\delta E_{\mathrm{int}}^{\mathrm{fc-CCSD}}(R)$ component. However, this conclusion changes substantially when moving to larger $R$. Indeed, near the equilibrium distance ($R=2.070$), the first three correlation components, i.e., $\delta E_{\mathrm{int}}^{\mathrm{fc-CCSD}}(R)$, $\delta E_{\mathrm{int}}^{\mathrm{fc-CCSDT}}(R)$, $\delta E_{\mathrm{int}}^{\mathrm{fc-CCSDTQ}}(R)$, bring a comparable contribution to the uncertainty. As a result, even if the first component can be improved substantially, this will not reduce the overall uncertainty to a comparable degree. We believe that it is unlikely that the terms  $\delta E_{\mathrm{int}}^{\mathrm{fc-CCSDT}}(R)$ and  $\delta E_{\mathrm{int}}^{\mathrm{fc-CCSDTQ}}(R)$ can be improved in near future in a meaningful way due to the steep scaling of the higher-order CC methods. Finally, when moving to a larger internuclear distance ($R=2.700$) the $\delta E_{\mathrm{int}}^{\mathrm{fc-CCSD}}(R)$ term no longer contributes significantly to the overall uncertainty. The uncertainties of the $\delta E_{\mathrm{int}}^{\mathrm{fc-CCSDT}}(R)$ and $\delta E_{\mathrm{int}}^{\mathrm{fc-CCSDTQ}}(R)$ components dominate here, but the uncertainty of $\delta E_{\mathrm{int}}^{\mathrm{fc-FCI}}(R)$ also becomes substantial. This is clearly the consequence of increasing multireference character of the molecule for larger $R$.

Moving on to the corrections beyond the non-relativistic BO approximation, the one-electron relativistic corrections $\delta E_{\mathrm{int}}^{\mathrm{rel1}}(R)$ are significant for all internuclear distances. The neglect of these corrections would require compensation through a substantial increase in the overall uncertainty estimate, by roughly 50\%, for all internuclear distances. Therefore, it is important to take this contribution into account also in the medium- and long-range of the potential. The other corrections bring a smaller contribution to the potential energy curve. The two-electron relativistic corrections, $\delta E_{\mathrm{int}}^{\mathrm{rel2}}(R)$, are anomalously large near the bottom of the potential as well as for smaller $R$, as discussed in Sec.~\ref{sec:rel12}. At $R=1.450$ and $R=2.070$, the term $\delta E_{\mathrm{int}}^{\mathrm{rel2}}(R)$ is only $2-3$ times smaller than $\delta E_{\mathrm{int}}^{\mathrm{rel1}}(R)$. However, the magnitude of the former diminishes quickly with increasing $R$. Already at $R=2.700$, is it about seven times smaller than $\delta E_{\mathrm{int}}^{\mathrm{rel1}}(R)$ and this trend continues for larger $R$. Therefore, we include the term $\delta E_{\mathrm{int}}^{\mathrm{rel2}}(R)$ in the short-range potential described here, but it can be safely neglected in the medium- and long-range. Finally, the remaining two corrections, $\delta E_{\mathrm{int}}^{\mathrm{QED}}(R)$ and $\delta E_{\mathrm{int}}^{\mathrm{DBOC}}(R)$, are very small for all internuclear distances and essentially drown in the combined uncertainties of other terms. Therefore, they are completely neglected in the short-range part of the potential and, consequently, also for larger $R$.

\section{Long-range potential energy curve}
\label{sec:lrpotential}

It is well known that when the internuclear distance in a diatomic molecule is sufficiently large, the potential energy curve can be represented by the asymptotic expansion:
\begin{align}
\label{cnasym}
    E_{\mathrm{int}}(R) = \sum_{n=6,8,10,\ldots} \frac{C_n}{R^n},
\end{align}
where $C_n$ are the dispersion coefficients. In Paper~I, we calculated the first three dispersion coefficients from first principles and estimated their uncertainties. The results are:
\begin{align}
    C_6 &= 23.956(51) \\
    C_8 &= 515.2(28) \\
    C_{10} &= 13733(73),
\end{align}
and hence they carry the relative uncertainty of about 0.21\%, 0.54\%, and 0.53\%, in the order of appearance.

We have to analyze what is the impact of the missing higher-order coefficients on the accuracy of the expansion and how large the value of $R$ has to be to make the above asymptotic expansion sufficiently accurate for the present purposes. Our goal at the moment is to obtain a rough estimate of the magnitude of these coefficients to be able to judge at which point the asymptotic expansion can be truncated. To this end, we first estimate values of the higher-order coefficients, $C_{12}$, $C_{14}$, and $C_{16}$ using the extrapolation formula introduced by Thakkar: \cite{thakkar88}
\begin{align}
\label{thakkar}
    C_{n+2} = \left( \frac{C_n}{C_{n-2}} \right)^3 C_{n-4}.
\end{align}
We begin with the impact of the $C_{12}$ coefficient. The term $C_{12}/R^{12}$ is still substantial---it contributes about 0.53\% and 0.34\% to the total interaction energy at $R=12.0$ and $R=13.0$, respectively. Clearly, this is not negligible from our point of view. The higher-order terms, by contrast, are significantly smaller. For example, $C_{14}/R^{14}$ contributes only 0.15\% and 0.082\% at $R=12.0$ and $R=13.0$, respectively, while the effect of $C_{16}/R^{16}$ is 0.054\% and 0.025\% at these points. Even higher-order terms are completely negligible. In total, truncation of the asymptotic expansion at the level of $C_{12}/R^{12}$, i.e., neglecting $C_{14}$, $C_{16}$, and higher-order coefficients, introduces an error no larger than 0.1\% in the interaction energy for $R\geq 13.0$. Taking into account that the lower-order coefficients carry a somewhat larger uncertainty, such level of truncation is acceptable. Therefore, we will use the asymptotic expansion~(\ref{cnasym}) truncated at $C_{12}$ to represent the potential energy curve for $R\geq 13.0$ further in the work.

While the coefficients $C_6$, $C_{8}$, $C_{10}$ are known from Paper~I, the value of $C_{12}$ has not been determined there, but it is necessary for sufficiently accurate representation of the potential at $R=13.0$ and beyond. Therefore, we calculate the value of $C_{12}$ using the formula in Eq.~(\ref{thakkar}). However, estimation of the uncertainty of $C_{12}$ obtained in this way is not straightforward. To get an insight into the possible magnitude of the error in $C_{12}$, we consider other systems for which the $C_{12}$ coefficient was calculated with a reliable first-principle method and can be used as a reference for the formula in Eq.~(\ref{thakkar}). The original paper of Thakkar \cite{thakkar88} used the hydrogen molecule as the main justification of the $C_n$ extrapolation. For this system, $C_{12}$ calculated from~(\ref{thakkar}) differs by about 1.4\% from the reference value. Next, for the ground state of the helium dimer, the $C_{12}$ coefficient was calculated explicitly in Ref.~\onlinecite{przybytek08} and it agrees with the extrapolated value to within 3.2\%. Finally, Patil and Tang \cite{patil97} calculated the $C_{12}$ asymptotic coefficients for lithium, sodium, potassium, and heavier alkali metals. The agreement between the calculated and extrapolated $C_{12}$ is, for example, about 8.6\% for lithium, 1.3\% for sodium, and 11.4\% for potassium, with a similar range of deviations for other alkali metals and their heteronuclear diatomics. Overall, this gives us confidence to conservatively assume that the $C_{12}$ coefficient for nitrogen obtained though extrapolation in Eq.~(\ref{thakkar}) is accurate to within 20\%. This leads to the estimate $C_{12} = 453\,715(90\,743)$.

The potential energy curve for $R\geq13.0$ is represented using the asymptotic expansion~(\ref{cnasym}), including terms up to $n=12$. We have to estimate the overall uncertainty of this representation which originates from two distinct sources. The first stems from the fact that the expansion is truncated at $n=12$. We previously estimated that this leads to the uncertainty no larger than 0.1\%. The second is the fact that the included asymptotic coefficients $C_6$, $C_8$, $C_{10}$, and $C_{12}$ carry an inherent uncertainty, as discussed above. However, after numerical experimentation we found that this uncertainty for $R\geq13.0$ is completely dominated by the uncertainty of the first coefficient, $C_6$. Even the large 20\% uncertainty of $C_{12}$ has essentially no impact on the accuracy of the expansion~(\ref{cnasym}). Therefore, we have to take into account only the estimated relative uncertainty of $C_6$, equal to 0.21\%, and the relative uncertainty originating from the truncation, 0.1\%. As these two sources of uncertainty are completely independent, we can apply the usual error propagation rules, and find that the total uncertainty of the long-range part of the potential is $\sqrt{(0.21\%)^2+(0.10\%)^2}=0.23\%$ for $R\geq13.0$. We adopt this estimate of the relative uncertainty for all internuclear distances above $R=13.0$. For the purposes of subsequent fitting of the total potential, the potential energy curve is evaluated according to Eq.~(\ref{cnasym}) for a grid of internuclear distances given in the Supplementary Material. \cite{supp}

\section{Medium-range potential energy curve}
\label{sec:mrpotential}

In the medium-range part of the potential, the multireference character of the electronic wavefunction becomes pronounced and the application of the single-reference CC theory is not practical. Therefore, in this regime we applied a genuine multireference method, namely the internally contracted multireference configuration-interaction (ic-MRCI) approach \cite{werner82,werner88,knowles88} in combination with the doubly augmented d-aug-cc-pV$X$Z, $X = 2,\ldots,6$, family of Gaussian basis sets. \cite{dunning89,kendall92,woon94,wilson96} In contrast with the short-range part of the potential, the use of doubly-augmented variants of the basis set becomes important for larger $R$ to accurately describe the molecular bond. Extrapolation to the CBS limit follows exactly the same protocol as in Sec.~\ref{sec:fc-ccsd}, so we skip the details here to avoid repetitions. For improved accuracy, the basis sets were fully uncontracted and the Davidson cluster correction with relaxed reference was applied. \cite{davidson77,knowles91,stark96} Similarly as in the case of the short-range potential, we take into account the leading-order one-electron relativistic corrections, $E_{\mathrm{rel1}}(R)$, that were evaluated perturbatively on top of the ic-MRCI wavefunction, analogously as in Sec.~\ref{sec:rel12}. The two-electron relativistic corrections were neglected in the medium-range part of the potential, as their influence appears to diminish quickly with $R$. Additionally, the QED and DBOC corrections were also neglected, following the approach used in the short-range potential. All electronic structure calculations reported in this section were performed using the {\sc Molpro} program package. \cite{werner12,werner20}

While ic-MRCI is a robust method with reasonable accuracy to cost ratio, it is not systematically improvable (at least within the implementations available). Therefore, it is not possible to apply a composite scheme, similar to those used for the short-range part of the potential, starting with the ic-MRCI results. In fact, only the basis set incompleteness error can be assessed systematically using the CBS extrapolation techniques and the random walk approach, but the same is not true for the uncertainties originating from the incomplete model of the ic-MRCI electronic wavefunction. Therefore, systematic estimation of the overall uncertainty is not possible based on the ic-MRCI results alone. To assign reliable uncertainty estimates to the ic-MRCI results, we calculate the relative errors with respect to the short-range component of the potential at $R=3.2$ (amounting to less than 1\%) and with respect to the long-range part of the potential at $R=13.0$ (amounting to about 15\%). We than assume the worst case scenario that the relative ic-MRCI errors grow linearly between these two points. This almost certainly leads to overestimation of the ic-MRCI uncertainties in the medium range, but enables to assign a conservative error bar to each of the ic-MRCI data points. Indeed, as ic-MRCI is not a size-consistent method, the relative error of this method (measured with respect to the exact result) is expected to monotonically increase as a function of $R$. Moreover, the accuracy of the method typically deteriorates at an increasing rate with $R$ which makes the relative error a convex function of $R$. If we join any two points on a convex curve with a line, the value of the function representing the curve will be below this line. Therefore, a linear interpolation of relative errors between two arbitrary points establishes an upper bound for the relative error between these points. The grid of internuclear distances for which ic-MRCI calculations were carried out is reported in the Supplementary Material. \cite{supp}

\section{Spectroscopic parameters}
\label{sec:spectro}

\subsection{Analytical fit of the potential}
\label{sec:fitpot}

In order to make the calculated results useful in determination of various spectroscopic properties of the N$_2$ molecule, an analytical representation of the potential energy curve must be obtained. After a considerable amount of numerical experimentation, we settled on the following form of the fitting function:
\begin{align}
\label{fitpot}
\begin{split}
    -E_{\mathrm{int}}(R) &= 
    e^{-\alpha_1 R}\sum_{n=-1}^{2} c_{n}^{(1)}\,R^n + 
    e^{-\alpha_2 R}\sum_{n=0}^{2} c_{n}^{(2)}\,R^n \\
    &+ e^{-\alpha_3 R^2}\sum_{n=0}^{2} c_{n}^{(3)}\,R^n
    - \sum_{n=3}^{6} \frac{C_{2n}}{R^{2n}}\,f_{2n}(\eta R),
\end{split}
\end{align}
where $f_n(x)$ are the Tang-Toennies damping functions defined as \cite{tang84}
\begin{align}
    f_n(x) = 1 - e^{-x}\sum_{k=0}^n \frac{x^k}{k!}.
\end{align}
These functions are introduced to remove the singularity of the long-range part in the $R\rightarrow 0$ limit. The parameter $\eta$ is adjustable and controls how strong the asymptotic tail is damped at short internuclear distances. The three exponential functions present in Eq.~(\ref{fitpot}) are responsible primarily for the reproduction of the short and medium part of the potential. Note that the coefficient $c_{-1}^{(1)}$ multiplies the term $e^{-\alpha R}/R$ which represents the Coulomb repulsion between nuclei as $R\rightarrow 0$. Therefore, in principle, $c_{-1}^{(1)}$ should be equal to the product of nuclear charges of the atomic nuclei. However, this form is valid only for very small values of $R$, while for the internuclear distances considered here, the nuclear charges are screened to a considerable degree by the core electrons. Therefore, we found that leaving $c_{-1}^{(1)}$ as an adjustable parameter considerably improves the quality of the fit. The asymptotic coefficients $C_n$ are not optimized and have fixed values given in Sec.~\ref{sec:lrpotential}. Therefore, the fitting function has, it total, four non-linear parameters ($\alpha_1$, $\alpha_2$, $\alpha_3$, $\eta$) and ten linear parameters. The total number of points in the potential energy curve used for fitting is equal to 82, see the Supplementary Material, \cite{supp} which is sufficiently large in comparison with the number of adjustable parameters in Eq.~(\ref{fitpot}) to avoid overfitting and unphysical oscillations.

The fitting is not applied directly to the calculated \emph{ab initio} interaction energies. Instead, in the fitting process we take into account that the theoretical results carry an inherent uncertainty. Let us denote the interaction energy calculated at the grid point $R_k$ by the symbol $E_k$ and the corresponding uncertainty at this point by $\delta E_k$. The fitting function was optimized to minimize the following quantity:
\begin{align}
    \tau = \sqrt{\frac{1}{N}\sum_k \left( \frac{E_k - E_{\mathrm{int}}(R_k)}{\delta E_k} \right)^2},
\end{align}
where $E_{\mathrm{int}}(R_k)$ is the value of the fitting function evaluated at the grid point $R_k$, and the summation over $k$ runs over all~82 aforementioned datapoints. According to this optimization target, the fitting function is adjusted to deviate from the calculated $\delta E_k$ by less than the inherent uncertainty $\delta E_k$ at each $R_k$.

The fitting process is performed in two steps. First, a large number $10^7$ of candidate values of the non-linear parameters $\alpha_1$, $\alpha_2$, $\alpha_3$, $\eta$ is randomly generated within the intervals (0,4) from the uniform distribution. For each candidate combination of $\alpha_1$, $\alpha_2$, $\alpha_3$, $\eta$, the linear parameters $c_n^{(i)}$ are determined using the standard least-squares procedure. From this initial pool, ten candidates are selected which give the lowest value of $\tau$. Finally, the $\alpha_1$, $\alpha_2$, $\alpha_3$, $\eta$ are fully optimized using the non-linear dogbox optimization method \cite{voglis2004rectangular} with the best ten candidates employed as the starting values.

In Table~\ref{tab:fit} we provide the final optimized parameters of the analytic representation of the potential energy curve. The Python implementation of this function enabling its evaluation for any $R$ is given in the Supplementary Material. \cite{supp} A plot illustrating the relative accuracy of the fit as a function of $R$ is also given in the Supplementary Material. \cite{supp} Note that the reported fitting function should not be used for internuclear distances smaller than $R=1.45$ which is the shortest $R$ for which the calculations were performed.

\begin{table}
\caption{\label{tab:fit}
Optimized linear and non-linear parameters of the fitting function~(\ref{fitpot}). The value of $\eta$ is 4.07718047. The large number of significant digits provided is necessary to reproduce the high accuracy of the fit in regions of the potential where a large numerical cancellation between terms in Eq.~(\ref{fitpot}) is observed.}
\begin{ruledtabular}
\begin{tabular}{clll}
 Param. & 
 \multicolumn{1}{c}{$i=1$} & 
 \multicolumn{1}{c}{$i=2$} & 
 \multicolumn{1}{c}{$i=3$} \\
 \hline\\[-1.2em]
 $c_{-1}^{(i)}$ & \phantom{$-$}169864.52928 & 
 \multicolumn{1}{c}{--} & \multicolumn{1}{c}{--} \\
 $c_{0}^{(i)}$  & $-$491715.748186 & 
 \phantom{$-$}345.450823 & 
 \phantom{$-$}1451.633626 \\
 $c_{1}^{(i)}$  & \phantom{$-$}520960.92251 & 
 $-$103.54848 & 
 \phantom{0}$-$807.374107 \\
 $c_{2}^{(i)}$  & $-$222581.529532 & 
 \phantom{$-$00}8.578901 & 
 \phantom{$-$0}168.827438 \\
 $\alpha_i$     & \phantom{$-$00000}4.77003203 & 
 \phantom{$-$00}1.92060894 & 
 \phantom{$-$000}0.74503316 \\
\end{tabular}
\end{ruledtabular}
\end{table}

\begin{table}[ht]
\caption{\label{tab:fitunc}
Optimized linear and non-linear parameters of the uncertainty fitting function, see Eq.~(\ref{fitunc}). The large number of significant digits provided is necessary to reproduce the high accuracy of the fit in regions of the potential where a large numerical cancellation between terms in Eq.~(\ref{fitunc}) is observed.}
\begin{ruledtabular}
\begin{tabular}{cd{8}d{7}d{8}}
 Param. & 
 \multicolumn{1}{c}{$i=1$} & 
 \multicolumn{1}{c}{$i=2$} & 
 \multicolumn{1}{c}{$i=3$} \\
 \hline\\[-1.2em]
 $c_{0}^{(i)}$  &  2.727034   & -6.191686  &  3.541105 \\
 $c_{1}^{(i)}$  & -1.146745   &  5.055429  & -2.36833\\
 $c_{2}^{(i)}$  &  0.130845   & -3.568555  &  0.474654\\
 $\alpha_i$     &  1.53112604 &  0.8944672 &  0.06594818 \\
\end{tabular}
\end{ruledtabular}
\end{table}

Finally, in the subsequent applications of the developed potential, we will require also a analytical representation of the uncertainty of $E_{\mathrm{int}}(R)$, denoted by the symbol $\delta E_{\mathrm{int}}(R)$. This quantity is not used in the present work, but will be critical in the determination of the properties of nitrogen which are relevant from the experimental point of view in the last paper of the series. To fit the potential's uncertainties determined at each internuclear distance according to the procedures described above, we employ the following function:
\begin{align}
\label{fitunc}
\begin{split}
    E_{\mathrm{unc}}(R) &= 
    e^{-\alpha_1 R}\sum_{n=0}^{2} c_{n}^{(1)}\,R^n + 
    e^{-\alpha_2 R^2}\sum_{n=0}^{2} c_{n}^{(2)}\,R^n \\
    &+ e^{-\alpha_3 R^4}\sum_{n=0}^{2} c_{n}^{(3)}\,R^n.
\end{split}
\end{align}
Optimized linear and non-linear parameters of the above function are reported in Table~\ref{tab:fitunc}.
The fit was obtained using the same procedure as for the potential energy curve, but using a smaller set of internuclear distances ($R\leq 20$) as the uncertainties are minuscule beyond this point. The average ratio of the uncertainty obtained from the fit to the actual uncertainty determined at the grid points is 1.03. This small increase of the overall uncertainty takes into account that the fitting function does not pass exactly through the \emph{ab initio} data points. We encountered a few points for which the fitted uncertainty is slightly smaller than the estimated theoretically (one point at $R=3.2$ and a few more points for distances $R\ge 12$).

\subsection{Rovibrational levels and spectroscopic constants}
\label{sec:rovibro}

The analytical representation of the potential energy curve obtained in the previous section can be used to determine the spectroscopic constants characterizing the electronic ground state of the N$_2$ molecule, as well as the rovibrational levels supported by the potential. We consider only the most naturally abundant (99.6\%) isotope of nitrogen, $^{14}$N, with the atomic mass $M(^{14}N) = 14.003\,074\,u$, according to Refs.~\onlinecite{Huang2021,Wang2021}.

The equilibrium bond length $R_e$ and the well-depth $D_e$ of the potential are found numerically by searching for the minimum of the analytic function, Eq.~(\ref{fitpot}). The harmonic vibrational frequency is found using the formula:
\begin{align}
    \omega_e^2 = \frac{1}{\mu} \left( \frac{\partial E_{\mathrm{int}}(R)}{\partial R^2} \right)\Big|_{R_e},
\end{align}
in atomic units, where $\mu$ is the reduced mass of the molecule. The second derivative of the potential was evaluated numerically using the fitting function~(\ref{fitpot}). The anharmonicity parameter $\omega_e x_e$ is found using a similar expression involving the third and fourth derivatives of the potential energy curve. The rotational constant $B_e$ is calculated using the rigid rotor approximation:
\begin{align}
    B_e = \frac{1}{2\mu R_e^2},
\end{align}
while the centrifugal distortion constant $d_e$ is evaluated using the non-rigid linear rotor model:
\begin{align}
    d_e = \frac{4B_e^2}{\omega_e^3}.
\end{align}
We use the lowercase letter for $d_e$ in contrast to $B_e$ to avoid confusion with the well-depth $D_e$. The first vibrational-rotational spectroscopic constant $\alpha_e$ (which is equal to the first mixed coefficient $Y_{11}$ in the Dunham expansion up to the sign change) is evaluated using the second-order perturbation theory expression:
\begin{align}
    \alpha_e = -\frac{6B_e^2}{\omega_e}\left[ 1 + \frac{R_e\, E'''_{\mathrm{int}}(R_e)}{3\mu\omega_e^2} \right],
\end{align}
where $E'''_{\mathrm{int}}(R_e)$ is the third derivative of the potential energy curve evaluated at the equilibrium distance $R_e$.

The rovibrational energy levels $E_{\nu J}$ of N$_2$ and the corresponding vibrational wavefunctions $|\chi_{\nu J}\rangle$ are found by solving one-dimensional radial Schr\"odinger equation for the relative nuclear motion; the renormalized Numerov-Cooley method \cite{cooley61,johnson1978renormalized} was used for this purpose. The dissociation energy $D_0$ of the molecule is found by adding to $D_e$ the energy of the ground state $(\nu=J=0)$ vibrational level, i.e., the zero-point vibrational energy (ZPE). In Table~\ref{tab:spectro} we report the results of the calculations of all aforementioned spectroscopic constants and compare them with the latest available experimental data. In the case where these parameters were not directly measured, the semi-empirical data of Le Roy \emph{et al.} \cite{le2006accurate} based on direct potential fit of the experimental rovibrational transition energies are used for the purposes of the comparison.

\begin{table}
\caption{\label{tab:spectro}
Comparison of spectroscopic constants obtained in this work and various literature data. All values except $R_e$, $D_e$ and $D_0$ are in $\mathrm{cm}^{-1}$.
}
\begin{ruledtabular}
\begin{tabular}{llll}
 Quantity & \multicolumn{1}{c}{This work} & \multicolumn{1}{c}{Experiment} &
 \multicolumn{1}{c}{Theor. work} \\
 \hline\\[-1.2em]
 $R_e(\mbox{\AA})$ & 1.0976 & 1.0977$^a$ & 1.0975$^f$\\
 $D_e(\mbox{eV})$ & 9.9034 & 9.900(1)$^b$ & 9.9050$^f$\\ 
 & & 9.899(1)$^c$ &\\ 
 & & 9.912(17)$^d$ &\\
 & & 9.905(5)$^e$ &\\
 $D_0(\mbox{eV})$ & 9.7575 & 9.7543(10)$^b$ & 9.7542(16)$^{g}$\\
 & & 9.7594(62)$^{d}$ &\\
 & & 9.759(5)$^e$ & \\
 $\omega_e$ & 2360.65 & 2358.53$^a$ & 2359.56$^f$ \\
 $\omega_e x_e$ & 14.259 & 14.3001$^a$ & 14.1093$^f$ \\
 $B_e$ & 1.9984 & 1.9983$^a$ & 1.99896$^f$ \\
 $a_e$ & 1.6946$\times10^{-2}$  & 1.7324$\times10^{-2}$$^{,a}$ &1.718$\times10^{-2}$$^{,f}$\\
 $d_e$ & $-5.73\times10^{-6}$& $-5.75\times10^{-6}$$^{,a}$ & \\
\end{tabular}
\end{ruledtabular}
\vspace{-0.3cm}
\begin{flushleft}
$^a$ From Ref.~\onlinecite{le2006accurate}\\
$^b$ From Ref.~\onlinecite{tang2005pulsed}, $D_e$ obtained in Ref.~\onlinecite{le2006accurate} from the same data\\
$^c$ From Ref.~\onlinecite{roncin1999vacuum}, $D_e$ obtained in Ref.~\onlinecite{le2006accurate} from the same data\\
$^d$ From Ref.~\onlinecite{buttenbender1934struktur}, $D_0$ obtained in Ref.~\onlinecite{chase1998nist} and $D_e$ in Ref.~\onlinecite{li2008full}\\
$^e$ From Ref.~\onlinecite{lofthus1977spectrum}\\
$^f$ From Ref.~\onlinecite{qin2019radiative}\\
$^g$ From Ref.~\onlinecite{thorpe21}
\end{flushleft}
\end{table}

First, regarding the equilibrium bond length, the agreement with the semi-empirical result is almost perfect, differing less than $10^{-4}$~\AA. In comparing the obtained well-depth $D_e$ with the experiment, we have to take into account the uncertainty of the calculations. It is reasonable to assume that the relative uncertainty of $D_e$ is essentially the same as for the interaction energy at the minimum of the potential (roughly 0.06\%). This leads to the absolute uncertainty of $6\,$~meV in the calculated value of $D_e$. If this uncertainty is taken into account, all experimental results lie within the estimated error bars of the theoretical calculations. In fact, this comparison also hints that the uncertainties of the potential energy curve estimated by us might have been too conservative near the bottom of the potential, because the actual deviation from the most recent experiment ($3.4$~meV) is roughly by a factor of two smaller than the estimated theoretical uncertainty.

\begin{table*}
\caption{\label{tab:rovib_rocin}
Comparison of energies (in cm$^{-1}$) of selected rovibrational transitions $E_{\nu',J'}- E_{00}$, i.e., $\nu''=J''=0$, obtained from our theoretical calculations with the experimental work of Roncin \emph{et al.} \cite{roncin1999vacuum} The symbol $\Delta$ denotes differences between the corresponding theoretical and experimental results, while $\Delta\%$ stand for the relative deviation between the theoretical and experimental data.
}
\begin{ruledtabular}
\begin{tabular}{r|rrrr|rrrr|rrrr}
   & \multicolumn{4}{c|}{$J'=0$} & \multicolumn{4}{c|}{$J'=10$} & \multicolumn{4}{c}{$J'=20$}  \\
  \hline
 $v'$  & \multicolumn{1}{c}{Exp.} & This work & \multicolumn{1}{c}{$\Delta$} & $\Delta\%$ 
       & \multicolumn{1}{c}{Exp.} & This work & \multicolumn{1}{c}{$\Delta$} & $\Delta\%$ 
       & \multicolumn{1}{c}{Exp.} & This work & \multicolumn{1}{c}{$\Delta$} & $\Delta\%$ \\
 \hline\\[-1.2em]
  0 &                &                 &            &            &                &                 &            &            &          834.7 &           834.7 &        0.1 &   0.01  \\
  1 &         2329.9 &          2332.0 &        2.1 &   0.09     &         2546.8 &          2549.0 &        2.2 &   0.09     &         3157.1 &          3159.6 &        2.4 &   0.08  \\
  2 &         4631.2 &          4635.2 &        4.0 &   0.09     &         4846.1 &          4850.3 &        4.2 &   0.09     &         5451.1 &          5455.5 &        4.5 &   0.08  \\
  3 &         6903.8 &          6909.5 &        5.7 &   0.08     &         7116.8 &          7122.6 &        5.8 &   0.08     &         7716.4 &          7722.5 &        6.1 &   0.08  \\
  4 &         9147.6 &          9154.6 &        7.0 &   0.08     &         9358.5 &          9365.8 &        7.3 &   0.08     &         9952.8 &          9960.3 &        7.5 &   0.08  \\
  5 &        11362.5 &         11370.7 &        8.1 &   0.07     &        11571.7 &         11579.9 &        8.2 &   0.07     &        12160.4 &         12168.9 &        8.5 &   0.07  \\
  6 &        13548.7 &         13557.5 &        8.8 &   0.06     &        13756.0 &         13764.8 &        8.9 &   0.06     &        14339.2 &         14348.4 &        9.1 &   0.06  \\
  7 &        15706.1 &         15715.2 &        9.1 &   0.06     &        15911.4 &         15920.5 &        9.1 &   0.06     &        16489.1 &         16498.5 &        9.4 &   0.06  \\
  8 &        17834.5 &         17843.6 &        9.1 &   0.05     &        18037.9 &         18047.0 &        9.1 &   0.05     &        18610.1 &         18619.4 &        9.3 &   0.05  \\
  9 &        19934.0 &         19942.8 &        8.8 &   0.04     &        20135.3 &         20144.2 &        8.9 &   0.04     &        20702.0 &         20711.1 &        9.1 &   0.04  \\
 10 &        22004.3 &         22012.8 &        8.5 &   0.04     &        22203.8 &         22212.3 &        8.4 &   0.04     &        22765.0 &         22773.6 &        8.6 &   0.04  \\
 11 &        24045.7 &         24053.7 &        8.0 &   0.03     &        24243.2 &         24251.2 &        8.0 &   0.03     &        24798.7 &         24806.9 &        8.2 &   0.03  \\
 12 &        26057.9 &         26065.5 &        7.6 &   0.03     &        26253.4 &         26261.0 &        7.5 &   0.03     &        26803.3 &         26811.1 &        7.8 &   0.03  \\
 13 &        28040.9 &         28048.2 &        7.3 &   0.03     &        28234.3 &         28241.7 &        7.4 &   0.03     &        28778.5 &         28786.2 &        7.6 &   0.03  \\
 14 &        29994.5 &         30001.8 &        7.3 &   0.02     &        30185.8 &         30193.3 &        7.5 &   0.02     &        30724.4 &         30732.1 &        7.7 &   0.03  \\
 15 &        31918.7 &         31926.3 &        7.6 &   0.02     &        32108.0 &         32115.8 &        7.8 &   0.02     &        32640.4 &         32649.0 &        8.6 &   0.03  \\
 16 &        33813.1 &         33821.8 &        8.7 &   0.03     &        34000.5 &         34009.2 &        8.8 &   0.03     &        34527.8 &         34536.7 &        8.9 &   0.03  \\
 17 &        35677.8 &         35688.1 &       10.3 &   0.03     &        35863.2 &         35873.5 &       10.3 &   0.03     &        36385.0 &         36395.3 &       10.2 &   0.03  \\
 18 &        37513.3 &         37525.2 &       11.9 &   0.03     &        37696.3 &         37708.6 &       12.3 &   0.03     &        38211.9 &         38224.6 &       12.7 &   0.03  \\
 19 &        39318.5 &         39333.0 &       14.5 &   0.04     &        39499.2 &         39514.3 &       15.2 &   0.04     &                &                 &            &         \\

\end{tabular}
\end{ruledtabular}
\end{table*}

A similar analysis holds also for $D_0$; its relative uncertainty is expected to be almost the same as $D_e$, because the ZPE contribution to $D_0$ is about 34 times smaller than $D_e$, so its uncertainty is essentially negligible. It is therefore safe to assume that the uncertainty of the theoretical value of $D_0$ reported in Table~\ref{tab:spectro} is about $6\,$meV. Taking this into account, the agreement with the experimental and semi-empirical values of $D_0$ is fully satisfactory and again hints at a significant overestimation of the theoretical uncertainty provided by us. Next, the harmonic vibrational frequency determined from the theoretical potential differs from the semi-empirical result of Le Roy \emph{et al.} \cite{le2006accurate} by about $2\!$~cm$^{-1}$ or about 0.09\% in relative terms, while the arhamonicity parameter $\omega_e x_e$ agrees to within roughly 0.3\%. The rotational constant $B_e$ differs from the semi-empirical data by only 0.005\%, but such near-perfect agreement might be accidental or a consequence of the fact that only the equilibrium distance $R_e$ is necessary to calculate $B_e$, and the latter parameter appears to be less sensitive to the quality of the potential energy curve. Somewhat larger deviations are observed for the centrifugal distortion constant $d_e$ and $\alpha_e$, but still within what can be expected taking the inherent uncertainty of the potential into account. It is also interesting to compare the value of $D_0$ of nitrogen molecule calculated by us with the best theoretical estimate available in the literature. Thorpe \emph{et al.} \cite{thorpe21} obtained $D_0=9.7542(16)$~eV using a composite CC scheme and also included minor corrections beyond the non-relativistic BO approximations. Their calculations were performed at only a single internuclear distance taken from the experiment and hence they are not directly comparable with our data, but nonetheless our results agree with Thorpe \emph{et al.} \cite{thorpe21} remarkably well if their respective uncertainties are taken into account and minor differences in the equilibrium bond length are ignored.

Next, we move to the analysis of the rovibrational levels supported by the developed theoretical potential energy curve, as this enables to compare results directly with the experiment without any additional assumptions. We focus on the two latest experimental sets of results for the nitrogen molecule. The first is the experimental work of Rocin \emph{et al.} \cite{roncin1999vacuum} who measured transition energies from the ground vibrational state $\nu''=J''=0$ to excited states with $v'\leq 19$ and a wide range of $J'$. As the total numbers of rovibrational levels reported in Ref.~\onlinecite{roncin1999vacuum} is too large to compare all of them, we selected $J'=0,10,20$ with the whole available range of $\nu'$ as representative examples. We checked that the conclusions found for this subset hold equally well for the remaining data from Ref.~\onlinecite{roncin1999vacuum}. In Table~\ref{tab:rovib_rocin} we provide a comparison of the selected transition energies from the experiment with theoretical results based on the potential energy curve reported in this work. Therein, we also include differences between the corresponding theoretical and experimental results ($\Delta$) and relative deviations between them ($\Delta\%$). One can see that with the increasing transition energy, the difference between the experiment and theory grows. However, if we take a look at relative differences, the trend is opposite -- relative errors smoothly decrease as a function of $\nu'$, from slightly less than $0.1\%$ for $\nu'=1$ to around $0.03-0.04\%$ for $\nu'=19$ (the highest vibrational level observed). Taking into consideration the inherent uncertainty of the \emph{ab initio} data, such level of agreement is better than expected based purely on the determined uncertainty of the theoretical results.

\begin{table*}
\caption{\label{tab:rovib_bendtsen}
Comparison of energies (in cm$^{-1}$) of selected rovibrational transitions $E_{0,J'}-E_{1,J''}$, i.e., $\nu''=1$, $\nu'=0$, obtained from our theoretical calculations with the experimental work of Bendtsen and Rasmussen, \cite{Bendtsen2000high} as taken from Le Roy \emph{et al.} \cite{le2006accurate} who assigned specific rovibrational quantum numbers to the experimental data. The symbol $\Delta$ denotes differences between the corresponding theoretical and experimental results. The relative deviation between the theoretical and experimental data are not shown, because they are almost constant along the whole dataset and amount to $0.09\%-0.1\%$.
}
\begin{ruledtabular}
\begin{tabular}{rr|ccc|rr|ccc}
 $J''$ & $J'$  & Exp. & This work & $\Delta$  & 
 $J''$ & $J'$  & Exp. & This work & $\Delta$  \\
 \hline\\[-1.2em]
    16 &    14    &     2448.4   &      2450.6   &     2.2 & 13 &     13 &        2326.8  &       2328.9 &       2.1 \\  
    15 &    13    &     2441.0   &      2443.2   &     2.2 & 14 &     14 &        2326.3  &       2328.4 &       2.2 \\
    14 &    12    &     2433.6   &      2435.8   &     2.2 & 15 &     15 &        2325.7  &       2327.9 &       2.2 \\
    13 &    11    &     2426.1   &      2428.3   &     2.2 & 16 &     16 &        2325.2  &       2327.4 &       2.2 \\
    12 &    10    &     2418.7   &      2420.8   &     2.2 & 17 &     17 &        2324.6  &       2326.8 &       2.2 \\
    11 &     9    &     2411.1   &      2413.3   &     2.1 & 18 &     18 &        2324.0  &       2326.2 &       2.2 \\
    10 &     8    &     2403.6   &      2405.7   &     2.1 & 19 &     19 &        2323.3  &       2325.5 &       2.2 \\
     9 &     7    &     2396.0   &      2398.1   &     2.1 & 20 &     20 &        2322.6  &       2324.8 &       2.2 \\
     8 &     6    &     2388.3   &      2390.5   &     2.1 & 21 &     21 &        2321.9  &       2324.1 &       2.2 \\
     7 &     5    &     2380.7   &      2382.8   &     2.1 & 22 &     22 &        2321.1  &       2323.4 &       2.2 \\
     6 &     4    &     2372.9   &      2375.1   &     2.1 & 23 &     23 &        2320.3  &       2322.6 &       2.2 \\
     5 &     3    &     2365.2   &      2367.3   &     2.1 & 24 &     24 &        2319.5  &       2321.7 &       2.3 \\
     4 &     2    &     2357.4   &      2359.5   &     2.1 &  0 &      2 &        2318.0  &       2320.1 &       2.1 \\
     3 &     1    &     2349.6   &      2351.7   &     2.1 &  1 &      3 &        2310.0  &       2312.1 &       2.1 \\
     2 &     0    &     2341.7   &      2343.8   &     2.1 &  2 &      4 &        2302.0  &       2304.1 &       2.1 \\
     0 &     0    &     2329.9   &      2332.0   &     2.1 &  3 &      5 &        2293.9  &       2296.0 &       2.1 \\
     1 &     1    &     2329.9   &      2332.0   &     2.1 &  4 &      6 &        2285.8  &       2287.9 &       2.1 \\
     2 &     2    &     2329.8   &      2331.9   &     2.1 &  5 &      7 &        2277.7  &       2279.8 &       2.1 \\
     3 &     3    &     2329.7   &      2331.8   &     2.1 &  6 &      8 &        2269.5  &       2271.6 &       2.1 \\
     4 &     4    &     2329.6   &      2331.7   &     2.1 &  7 &      9 &        2261.3  &       2263.4 &       2.1 \\
     5 &     5    &     2329.4   &      2331.5   &     2.1 &  8 &     10 &        2253.1  &       2255.2 &       2.1 \\
     6 &     6    &     2329.2   &      2331.3   &     2.1 &  9 &     11 &        2244.8  &       2247.0 &       2.1 \\
     7 &     7    &     2328.9   &      2331.1   &     2.1 & 10 &     12 &        2236.6  &       2238.7 &       2.1 \\
     8 &     8    &     2328.7   &      2330.8   &     2.1 & 11 &     13 &        2228.2  &       2230.3 &       2.1 \\
     9 &     9    &     2328.3   &      2330.5   &     2.1 & 12 &     14 &        2219.9  &       2222.0 &       2.1 \\
    10 &    10    &     2328.0   &      2330.1   &     2.1 & 13 &     15 &        2211.5  &       2213.6 &       2.1 \\
    11 &    11    &     2327.6   &      2329.8   &     2.1 & 14 &     16 &        2203.1  &       2205.2 &       2.1 \\
    12 &    12    &     2327.2   &      2329.3   &     2.1 & 16 &     18 &        2186.2  &       2188.3 &       2.2 \\
\end{tabular}
\end{ruledtabular}
\end{table*}

It is also interesting to compare the errors in the vibrational levels obtained by us with other theoretical results available in the literature. A broad overview of the literature up to year 2018 devoted to the theoretical calculation of the interaction energies of nitrogen molecule is given in the paper of Qin \emph{et al.} \cite{qin2019radiative} However, many of the relevant papers were not concentrated on the ground state or the accuracy of the results was not the primary focus. A notable exception is the 2008 paper of Li and Paldus \cite{li2008full} where the $J=0$ vibrational levels were calculated and compared with the experiment. For the first few vibrational levels, accuracy of their data is slightly better than ours, but becomes roughly order of magnitude less accurate for higher $\nu$. More recently, a similar set of data was published by Ding \emph{et al.} \cite{ding2023collision} but their work was not focused primarily on the ground state and their accuracy is inferior to ours in comparison with the experiment. Quite recently, Hammami \emph{et al.} \cite{hammami2026potential} carried a study of a large number of electronic states of the nitrogen molecule, including the ground state. The accuracy of the $J=0$ vibrational transitions reported by them is somewhat better in comparison with ours for $\nu<10$, but becomes progressively worse for $\nu>10$ in relation to the measurement. Moreover, it has to be pointed out that the relativistic and core-correlation effects were not taken into account in Ref.~\onlinecite{hammami2026potential}. As their combined magnitude is substantial, see Table~\ref{tab:short}, their inclusion would likely decrease the overall agreement with the experimental data.

The next comparison is with the experimental data of Bendtsen and Rasmussen \cite{Bendtsen2000high} who measured energies of rovibrational transitions $E_{0,J'}-E_{1,J''}$, i.e., $\nu''=1$, $\nu'=0$ for a subset of rotational quantum numbers $J''$ and $J'$. In Table~\ref{tab:rovib_bendtsen} we provide a comparison between these experimental and our theoretical results and differences ($\Delta$) between them. We do not report separately the relative deviations in this case, because they are almost constant for the whole data set and amount to $0.09\%-0.1\%$. Similarly as before, this level of discrepancy is consistent with the expected uncertainty of the theoretical potential energy curve.

\begin{figure}
    \centering
    \includegraphics[scale=0.85]{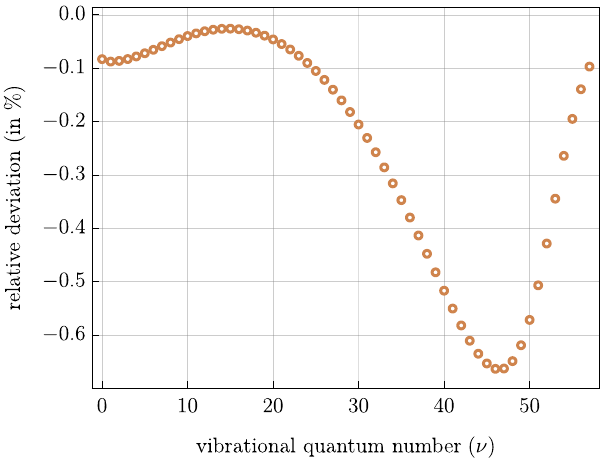}
    \caption{Relative deviation (in percent) between the vibrational energy levels $E_{\nu 0}$ ($J=0$) obtained using the theoretical potential energy curve developed in this work and the data of Le Roy \emph{et al.} \cite{le2006accurate}}
    \label{fig:leroy}
\end{figure}

Finally, we compare energies of vibrational levels with the work of Le Roy \emph{et al.} \cite{le2006accurate} for states with $J=0$ and $\nu>19$ which have never been observed experimentally through direct measurements. The semi-empirical potential developed by Le Roy \emph{et al.} \cite{le2006accurate} was obtained through direct potential fit of the observed transition energies using a model potential energy curve based on the Morse/long-range analytical form. In Fig.~\ref{fig:leroy} we present relative deviations of the calculated vibrational energy levels from our potential and based on the data of Le Roy \emph{et al.} \cite{le2006accurate} One can see that up to $\nu\approx 25$ the differences do not exceed $0.1\%$ which is consistent with the estimated uncertainty of the theoretical potential in the vicinity of the equilibrium internuclear distance. However, moving to higher $\nu$ the differences increase and reach the level of $0.6-0.7\%$ for $\nu\approx 45$. 
This behavior is not unexpected as the long-range component of the theoretical potential is around three times less accurate (in relative terms) than near the minimum. Moreover, the medium range of the potential can be even somewhat less accurate. Finally, in contrast to the potential of Le Roy \emph{et al.}, \cite{le2006accurate} we were unable to locate the last two bound states observed by them, namely $\nu = 59$ and $\nu=60$, using the potential described in this work. Nonetheless, a satisfactory accuracy level of around $0.1\%$ is preserved for vibrational levels with energies as high as $\approx 50\,000\!$~cm$^{-1}$. This is entirely sufficient from the point of view of applications in metrology, as vibrational levels higher than that would only be relevant for extremely high temperatures which are beyond the range of applicability of the current gas metrology techniques.

\section{Conclusions}
\label{sec:conclusions}

In this work, we have developed an accurate, first-principles potential energy curve for the electronic ground state of the nitrogen molecule. Because the dissociation of N$_2$ involves breaking a strong triple bond, the electronic wavefunction acquires a pronounced multireference character at larger internuclear separations. To overcome this challenge, we partitioned the potential energy curve into three distinct regions. For the short-range region, which dictates the bulk of the temperature-dependent thermophysical properties relevant to gas metrology, we employed a composite scheme. This protocol is rooted in a hierarchy of coupled-cluster methods up to the CCSDTQ(P) level, augmented by full configuration interaction (FCI) estimates, core-core and core-valence correlation, and complete basis set extrapolations. We also explicitly evaluated higher-order physical effects, demonstrating that while one- and two-electron relativistic corrections are necessary for a balanced representation, the QED and DBOC corrections remain negligible at the current target accuracy. Crucially, we propagated the uncertainties originating from basis set incompleteness and method truncation throughout the entire scheme, providing a rigorous and systematic error budget for the \emph{ab initio} points. The medium-range potential was computed using the internally contracted multireference configuration-interaction (ic-MRCI) method to properly handle the bond-breaking regime. The long-range tail was modeled using a mathematically robust asymptotic expansion of the dispersion coefficients: $C_6$ through $C_{10}$ calculated in Paper~I, and $C_{12}$ estimated in this work.

The quality of the developed potential was validated by computing the spectroscopic constants (including $R_e$, $D_e$, $\omega_e$, $B_e$) and the rovibrational energy levels of the $^{14}$N$_2$ isotopologue. The theoretical predictions exhibit good agreement with the most recent experimental and semi-empirical literature data. Relative deviations in transition energies consistently fall below $0.1\%$, and in all cases the residual differences between theory and experiment are significantly smaller than theoretical uncertainty estimates. In the forthcoming part of the series, this potential will be utilized to determine the temperature dependence of fundamental properties of nitrogen gas, such as polarizability and magnetic susceptibility, bridging the gap between rigorous \emph{ab initio} theory and precision thermometry and pressure metrology experiments.

\section*{Supplementary Material}

See the Supplementary Material for the raw results of all calculations reported in this work, summary of the results for the whole grid of internuclear distances, complete comparison with the results of Le Roy \emph{et al.}, \cite{le2006accurate} and Python implementation of the fitting functions for the potential and its uncertainty.

\begin{acknowledgments}
The project (22IEM04 MQB-Pascal) has received funding from the European Partnership on Metrology, co-financed from the European Union’s Horizon Europe Research and Innovation Programme and by the Participating States. We gratefully acknowledge Poland's high-performance Infrastructure PLGrid (HPC Centers: ACK Cyfronet AGH, PCSS, CI TASK, WCSS) for providing computer facilities and support within computational grant PLG/2025/018692. The authors also thank Pozna\'n Supercomputing and Networking Center for the computational grant pl0458-01. 
\end{acknowledgments}

\section*{Data Availability Statement}

The data that support the findings of this study are available within the article and its supplementary material, and openly available in Zenodo repository at \url{https://doi.org/10.5281/zenodo.22662098}, reference number 22662098.

\bibliography{nitrogen-paper2}

\end{document}